\documentclass[prd,onecolumn,nofootinbib]{revtex4}
\usepackage{bm,amsmath,amssymb,graphicx}

\begin{document}

\title{Gravitational Collapse with Torsion and Universe in a Black Hole}
\author{Nikodem Pop{\l}awski}
\altaffiliation{NPoplawski@newhaven.edu, Department of Mathematics and Physics, University of New Haven, 300 Boston Post Road, West Haven, CT 06516, USA}

\begin{abstract}
We consider gravitational collapse of a sphere of a fluid with torsion generated by spin, which forms a black hole.
We use the Tolman metric and the Einstein--Cartan field equations with a relativistic spin fluid as a source.
We show that gravitational repulsion of torsion prevents a singularity, replacing it with a nonsingular bounce.
Quantum particle creation during contraction prevents shear from overcoming torsion.
Particle creation during expansion can generate a finite period of inflation and produce large amounts of matter.
The resulting closed universe on the other side of the event horizon may have several bounces.
Such a universe is oscillatory, with each cycle larger than the preceding cycle, until it reaches a size at which dark energy dominates and expands indefinitely.
Our universe might have therefore originated from a black hole existing in another universe.\\

\noindent
In: {\em Regular Black Holes: Towards a New Paradigm of Gravitational Collapse}, Cosimo Bambi (ed.), chapter 13, pp. 485--499 (Springer, Singapore, 2023).
\end{abstract}
\maketitle

\section{Torsion and regular black holes}
The general theory of relativity (GR) is a geometric theory of gravitation, in which curved spacetime is described by the metric tensor.
The affine connection, which describes differentiation in a curved spacetime, is constrained to be symmetric and given by the Christoffel symbols constructed from the metric tensor and its derivatives \cite{GR2,GR1,LL2}.
In the presence of a symmetric connection, the orbital angular momentum of a free particle is conserved \cite{Lord,Niko}.
However, the Dirac equation of relativistic quantum mechanics predicts the conservation law for the total (orbital plus spin) angular momentum of a free Dirac particle, allowing a spin-orbit interaction \cite{Dirac}.
This conservation law arises naturally in theories of gravity with an asymmetric affine connection \cite{req}.
The antisymmetric part of the connection is referred to as the torsion tensor \cite{Schr1,Schr3,Schr2}.

The simplest theory of gravity that extends GR by relaxing the symmetry constraint of the affine connection and allowing torsion is the Einstein--Cartan (EC) theory \cite{EC6,EC7,EC5,EC8,EC2,Lord,Niko,non,EC9,EC10,EC1,EC3,EC4}.
In this theory, expanded by Sciama and Kibble, the Lagrangian density for the gravitational field is the same as in GR: it is proportional to the Ricci scalar constructed from the connection 
The field equations, obtained from varying the action for gravity and matter with respect to the torsion tensor, determine the torsion tensor to be proportional to the spin tensor of fermionic matter \cite{EC8,EC2,Lord,Niko,EC4}.
Consequently, EC can be rewritten as GR with the symmetric Levi-Civita connection, in which the energy--momentum tensor of matter acquires additional terms that are quadratic in the spin tensor.
In vacuum, torsion vanishes and EC reduces to GR.
Following the multipole expansion of the conservation law for the spin tensor in EC, fermionic matter has a form of a spin fluid: an ideal fluid with a spin tensor that is linear in the matter four-velocity \cite{NSH}.

Hehl \cite{Hehl}, Trautman \cite{Tra1,Tra2,Tra3}, and Kopczy\'{n}ski \cite{Kop1,Kop2} discovered that torsion can generate gravitational repulsion and prevent the formation of a cosmological singularity in a homogeneous and isotropic universe described by the Friedmann--Lema\^{i}tre--Robertson--Walker (FLRW) metric \cite{GR2,GR1,FLRW1,LL2,FLRW2,Lord,FLRW3,FLRW4} when spins of fermions are aligned.
A singularity can be also avoided for randomly oriented spins because macroscopic averaging of the quadratic spin terms in the energy--momentum tensor gives nonzero values \cite{HHK}.
The effective energy density $\tilde{\epsilon}$ and pressure $\tilde{p}$ of a spin fluid with randomly oriented spins are given by
\begin{equation}
    \tilde{\epsilon}=\epsilon-\alpha n_\textrm{f}^2,\quad\tilde{p}=p-\alpha n_\textrm{f}^2,
    \label{intro1}
\end{equation}
where $\epsilon$ is the thermodynamic energy density, $p$ is its pressure, $n_\textrm{f}$ is the number density of fermions, and $\alpha=\kappa(\hbar c)^2/32$ \cite{HHK,NP,ApJ1,ApJ2,Gabe} with $\kappa=8\pi G/c^4$.
At densities much lower than nuclear density, the effects of torsion can be neglected and EC is indistinguishable from GR, passing all observational tests.
At densities much higher than nuclear density, that exist in black holes or in the very early universe, the negative corrections from the spin-torsion coupling in (\ref{intro1}) violate the strong energy condition and manifest themselves as repulsive gravity that may prevent the formation of singularities \cite{iso2,iso1,iso3,cosmo,iso4,ApJ1,ApJ2}.

In the presence of torsion, the collapsing matter in a black hole would avoid a singularity and instead reach a nonsingular bounce, after which it would expand as a new, closed universe \cite{ApJ1,ApJ2,Gabe} whose total energy is zero \cite{energy1,energy2}.
A nonsingular bounce can also occur if the spin tensor is completely antisymmetric \cite{spi3,spi2,spi1}.
If a black hole creates a baby universe on the other side of its event horizon, then such a universe would be connected to the parent universe through an Einstein--Rosen bridge (wormhole) \cite{Ein,ER}.
The formation and subsequent dynamics of such a universe cannot be observed outside the black hole because of the infinite redshift at the horizon.
Consequently, if our universe is closed \cite{closed1,closed2}, then it might have originated as a baby universe after a bounce in the interior of a parent black hole existing in another universe \cite{Pat8,Pat3,Pat4,Pat6,Pat1,Pat2,ER,cosmo,ApJ1,ApJ2,collapse1,collapse2,Pat5,Pat9,Pat7}.

\section{Einstein--Cartan gravity}
The EC theory of gravity naturally extends GR to include matter with spin, providing a more complete account of local gauge invariance with respect to the Poincar\'{e} group \cite{EC8,EC2,Niko,EC4}.
In this theory, the affine connection $\Gamma^{\,\,k}_{i\,j}$ has an antisymmetric part: the torsion tensor
\begin{equation}
    S^k_{\phantom{k}ij}=\frac{1}{2}(\Gamma^{\,\,k}_{i\,j}-\Gamma^{\,\,k}_{j\,i}).
\end{equation}
The curvature tensor is given by $R^i_{\phantom{i}mjk}=\partial_j \Gamma^{\,\,i}_{m\,k}-\partial_k \Gamma^{\,\,i}_{m\,j}+\Gamma^{\,\,i}_{l\,j}\Gamma^{\,\,l}_{m\,k}-\Gamma^{\,\,i}_{l\,k}\Gamma^{\,\,l}_{m\,j}$ and its contraction gives the Ricci tensor $R_{ik}=R^j_{\phantom{j}ijk}$.
The metricity condition $\nabla_j g_{ik}=0$, where $\nabla_i$ denotes the covariant derivative for the connection $\Gamma^{\,\,k}_{i\,j}$, gives
\begin{equation}
    \Gamma^{\,\,k}_{i\,j}=\{^{\,\,k}_{i\,j}\}+C^k_{\phantom{k}ij},
\end{equation}
where $\{^{\,\,k}_{i\,j}\}=(1/2)g^{km}(\partial_j g_{mi}+\partial_i g_{mj}-\partial_m g_{ij})$ are the Christoffel symbols for the metric tensor $g_{ik}$ and $C^k_{\phantom{k}ij}=S^k_{\phantom{k}ij}+S_{ij}^{\phantom{ij}k}+S_{ji}^{\phantom{ji}k}$ is the contortion tensor.
The curvature tensor can be decomposed as $R^i_{\phantom{i}klm}=P^i_{\phantom{i}klm}+D_l C^i_{\phantom{i}km}-D_m C^i_{\phantom{i}kl}+C^j_{\phantom{j}km}C^i_{\phantom{i}jl}-C^j_{\phantom{j}kl}C^i_{\phantom{i}jm}$, where $P^i_{\phantom{i}klm}$ is the Riemann tensor (the curvature tensor constructed from the Levi-Civita connection $\{^{\,\,k}_{i\,j}\}$) and $D_i$ denotes the covariant derivative for the connection $\{^{\,\,k}_{i\,j}\}$ \cite{Niko}.

The EC gravity is based on the Lagrangian density of the gravitational field that is proportional to the Ricci curvature scalar $R=R_{ik}g^{ik}$, similarly to GR which uses the curvature scalar constructed from $\{^{\,\,k}_{i\,j}\}$.
The field equations are obtained from varying the total action for the gravitational field and matter, $(1/c)\int[-(1/2\kappa)R\sqrt{-g}+\mathcal{L}_\textrm{m}]d^4x$, where $\mathcal{L}_\textrm{m}$ is the Lagrangian density for matter, and $g=\textrm{det}(g_{ik})$, with respect to the metric and torsion tensors.
Varying the action with respect to the torsion tensor gives the Cartan field equations, which relate algebraically the torsion tensor to the canonical spin tensor of matter $s^{ijk}=2(\delta\mathcal{L}_\textrm{m}/\delta C_{ijk})/\sqrt{-g}$ \cite{EC8,EC2,Lord,Niko,EC4}:
\begin{equation}
    S_{jik}-S_i g_{jk}+S_k g_{ji}=-\frac{1}{2}\kappa s_{ikj},
    \label{field1}
\end{equation}
where $S_i=S^k_{\phantom{k}ik}$.
Varying the action with respect to the metric tensor gives the Einstein field equations, which relate the curvature tensor to the canonical energy--momentum tensor of matter $\sigma_{ik}$:
\begin{equation}
    R_{ik}-\frac{1}{2}Rg_{ik}=\kappa\sigma_{ki}.
    \label{field2}
\end{equation}
The canonical energy--momentum tensor is related to the metric energy--momentum tensor $T_{ik}=2(\delta\mathcal{L}_\textrm{m}/\delta g^{ik})/\sqrt{-g}$ by $T_{ik}=\sigma_{ik}-(1/2)(\nabla_j-2S_j)(s_{ik}^{\phantom{ik}j}-s_{k\phantom{j}i}^{\phantom{k}j}+s^j_{\phantom{j}ik})$ \cite{EC8,Lord,Niko}.

The field equations (\ref{field1}) and (\ref{field2}) can be combined to give
\begin{equation}
    G^{ik}=\kappa T^{ik}+\frac{1}{2}\kappa^2\biggl(s^{ij}_{\phantom{ij}j}s^{kl}_{\phantom{kl}l}-s^{ij}_{\phantom{ij}l}s^{kl}_{\phantom{kl}j}-s^{ijl}s^k_{\phantom{k}jl}+\frac{1}{2}s^{jli}s_{jl}^{\phantom{jl}k}+\frac{1}{4}g^{ik}(2s^{\phantom{j}l}_{j\phantom{l}m}s^{jm}_{\phantom{jm}l}-2s^{\phantom{j}l}_{j\phantom{l}l}s^{jm}_{\phantom{jm}m}+s^{jlm}s_{jlm})\biggr),
    \label{field3}
\end{equation}
where $G_{ik}=P_{ik}-(1/2)Pg_{ik}$ is the Einstein tensor of general relativity constructed from the contractions of the Riemann tensor, $P_{ik}=P^j_{\phantom{j}ijk}$ and $P=P_{ik}g^{ik}$ \cite{EC8,Niko}.
The second term on the right-hand side of (\ref{field3}) is a correction to curvature from spin.
The spin tensor also appears in $T_{ik}$ because $\mathcal{L}_\textrm{m}$ depends on torsion.

Quarks and leptons, that compose all stars, are fermions described in relativistic quantum mechanics by the Dirac equation.
Since Dirac fields couple minimally to the torsion tensor, torsion at microscopic scales is generated in the presence of fermions \cite{EC8,EC2,Niko,EC4}.
At macroscopic scales, such particles can be averaged and described as a spin fluid \cite{iso2,HHK,iso1,cosmo}.
If particles have randomly oriented spins, then the macroscopic spacetime averages of the spin and of the spin gradients vanish.
However, the spin terms in the Einstein equations (\ref{field3}) are quadratic in the spin tensor and do not vanish after averaging \cite{HHK}.
These contributions from spin to the field equations are significant only at densities of matter much larger than the density of nuclear matter (on the order of the Cartan density \cite{non}) because of the factor $\kappa^2$.
At nuclear densities and below, the predictions of EC do not differ from the predictions of the metric general relativity, and reduce to them in vacuum, where torsion vanishes.

The Bianchi identity $\nabla_{[l}R^i_{\phantom{i}|n|jk]}=2R^i_{\phantom{i}nm[j}S^m_{\phantom{m}kl]}$ and the cyclic identity $R^m_{\phantom{m}[jkl]}=-2\nabla_{[l}S^m_{\phantom{m}jk]}+4S^m_{\phantom{m}n[j}S^n_{\phantom{n}kl]}$, together with the Einstein
and Cartan field equations give the conservation laws for the canonical energy--momentum and spin tensors: $D_j \sigma^{ij}=C_{jk}^{\phantom{jk}i}\sigma^{jk}+(1/2)s_{klj}R^{klji}$ and $\nabla_k s_{ij}^{\phantom{ij}k}-2S_k s_{ij}^{\phantom{ij}k}=\sigma_{ij}-\sigma_{ji}$ \cite{EC8,Niko}.
Using the multipole expansion for these laws leads to the formulas for the macroscopic canonical energy--momentum and spin tensors in the point-particle approximation \cite{NSH,Niko}.
These tensors define a spin fluid \cite{HHK}.
The canonical energy--momentum tensor of a spin fluid is given by
\begin{equation}
    \sigma_{ij}=c\Pi_i u_j-p(g_{ij}-u_i u_j),
\end{equation}
and its canonical spin tensor is given by
\begin{equation}
    s_{ij}^{\phantom{ij}k}=s_{ij}u^k,\quad s_{ij}u^j=0,\quad s^2=\frac{1}{2}s_{ij}s^{ij}>0,
\end{equation}
where $\Pi_i$ is the four-momentum density of the fluid, $u^i$ is its four-velocity, and $s_{ij}$ is its spin density.

The field equations (\ref{field3}) give \cite{HHK}
\begin{equation}
    G^{ij}=\kappa\Bigl(\epsilon-\frac{1}{4}\kappa s^2\Bigr)u^i u^j-\kappa\Bigl(p-\frac{1}{4}\kappa s^2\Bigr)(g^{ij}-u^i u^j)-\frac{1}{2}\kappa(\delta^l_k+u_k u^l)D_l(s^{ki}u^j+s^{kj}u^i),
    \label{field4}
\end{equation}
where $\epsilon=c\Pi_i u^i$.
For randomly oriented spins of particles, the last term on the right-hand side of (\ref{field4}) vanishes after averaging.
Thus the Einstein--Cartan equations for such a spin fluid are equivalent to the general-relativistic Einstein equations for an ideal fluid with the effective energy density $\tilde{\epsilon}=\epsilon-\kappa s^2/4$ and pressure $\tilde{p}=p-\kappa s^2/4$ \cite{HHK,EC8,Niko}.
The square of the spin density for a fluid consisting of fermions with random spin orientation is given by \cite{NP}
\begin{equation}
s^2=\frac{1}{8}(\hbar c n_\textrm{f})^2.
\end{equation}
Consequently, the effective energy density and pressure of a spin fluid give (\ref{intro1}).

\section{Gravitational collapse of a homogeneous sphere}
In this Chapter, we consider gravitational collapse of a sphere of a homogeneous spin fluid that is initially at rest, to demonstrate the formation of a new universe in a black hole \cite{collapse1,collapse2}.
The presented work extends the analysis of collapse of a dustlike sphere by Landau and Lifshitz \cite{LL2}, based on the work of Tolman \cite{Tol} and Oppenheimer and Snyder \cite{OS}.
This formalism relates the initial scale factor of the universe in a black hole to the initial radius and mass of the black hole.
In the absence of pressure gradients, such a collapse can be described in a system of coordinates that is both synchronous and comoving \cite{LL2}.

For a spherically symmetric gravitational field in spacetime filled with an ideal fluid, the geometry is given by the Tolman metric \cite{LL2,Tol}:
\begin{equation}
    ds^2=e^{\nu(\tau,R)}c^2 d\tau^2-e^{\lambda(\tau,R)}dR^2-e^{\mu(\tau,R)}(d\theta^2+\mbox{sin}^2\theta\,d\phi^2),
    \label{grav1}
\end{equation}
where $\nu$, $\lambda$ and $\mu$ are functions of a time coordinate $\tau$ and a radial coordinate $R$.
We can still apply coordinate transformations $\tau\rightarrow \tau'(\tau)$ and $R\rightarrow R'(R)$ without changing the form of the metric (\ref{grav1}).
The components of the Einstein tensor corresponding to (\ref{grav1}) that do not vanish identically are \cite{LL2,Tol}:
\begin{eqnarray}
& & G_0^0=-e^{-\lambda}\Bigl(\mu''+\frac{3\mu'^2}{4}-\frac{\mu'\lambda'}{2}\Bigr)+\frac{e^{-\nu}}{2}\Bigl(\dot{\lambda}\dot{\mu}+\frac{\dot{\mu}^2}{2}\Bigr)+e^{-\mu}, \nonumber \\
& & G_1^1=-\frac{e^{-\lambda}}{2}\Bigl(\frac{\mu'^2}{2}+\mu'\nu'\Bigr)+e^{-\nu}\Bigl(\ddot{\mu}-\frac{\dot{\mu}\dot{\nu}}{2}+\frac{3\dot{\mu}^2}{4}\Bigr)+e^{-\mu}, \nonumber \\
& & G_2^2=G_3^3=-\frac{e^{-\nu}}{4}(\dot{\lambda}\dot{\nu}+\dot{\mu}\dot{\nu}-\dot{\lambda}\dot{\mu}-2\ddot{\lambda}-\dot{\lambda}^2-2\ddot{\mu}-\dot{\mu}^2) \nonumber \\
& & -\frac{e^{-\lambda}}{4}(2\nu''+\nu'^2+2\mu''+\mu'^2-\mu'\lambda'-\nu'\lambda'+\mu'\nu'), \nonumber \\
& & G_0^1=\frac{e^{-\lambda}}{2}(2\dot{\mu}'+\dot{\mu}\mu'-\dot{\lambda}\mu'-\dot{\mu}\nu'),
\label{grav2}
\end{eqnarray}
where a dot denotes differentiation with respect to $c\tau$ and a prime denotes differentiation with respect to $R$.

In a comoving frame of reference, the spatial components of the four-velocity $u^i$ vanish.
Accordingly, the nonzero components of the energy--momentum tensor for a spin fluid, $T_{ik}=(\tilde{\epsilon}+\tilde{p})u_i u_k-\tilde{p}g_{ik}$, are: $T^0_0=\tilde{\epsilon}$, $T^1_1=T^2_2=T^3_3=-\tilde{p}$.
The Einstein field equations $G^i_k=\kappa T^i_k$ in this frame of reference are:
\begin{equation}
    G_0^0=\kappa\tilde{\epsilon},\quad G_1^1=G_2^2=G_3^3=-\kappa\tilde{p},\quad G_0^1=0.
\end{equation}
The covariant conservation of the energy--momentum tensor gives
\begin{equation}
    \dot{\lambda}+2\dot{\mu}=-\frac{2\dot{\tilde{\epsilon}}}{\tilde{\epsilon}+\tilde{p}},\,\,\,\nu'=-\frac{2\tilde{p}'}{\tilde{\epsilon}+\tilde{p}},
    \label{grav4}
\end{equation}
where the constants of integration depend on the allowed transformations $\tau\rightarrow \tau'(\tau)$ and $R\rightarrow R'(R)$.

If the pressure is homogeneous (no pressure gradients), then $p'=0$ and $p=p(\tau)$.
In this case, the second equation in (\ref{grav4}) gives $\nu'=0$.
Therefore, $\nu=\nu(\tau)$ and a transformation $\tau\rightarrow \tau'(\tau)$ can bring $\nu$ to zero and $g_{00}=e^\nu$ to 1.
The system of coordinates becomes synchronous \cite{LL2}.
Defining $r(\tau,R)=e^{\mu/2}$ turns (\ref{grav1}) into
\begin{equation}
    ds^2=c^2 d\tau^2-e^{\lambda(\tau,R)}dR^2-r^2(\tau,R)(d\theta^2+\mbox{sin}^2\theta\,d\phi^2).
    \label{grav5}
\end{equation}
The Einstein equations (\ref{grav2}) reduce to
\begin{eqnarray}
& & \kappa\tilde{\epsilon}=-\frac{e^{-\lambda}}{r^2}(2rr''+r'^2-rr'\lambda')+\frac{1}{r^2}(r\dot{r}\dot{\lambda}+\dot{r}^2+1), \nonumber \\
& & -\kappa\tilde{p}=\frac{1}{r^2}(-e^{-\lambda}r'^2+2r\ddot{r}+\dot{r}^2+1), \nonumber \\
& & -2\kappa\tilde{p}=-\frac{e^{-\lambda}}{r}(2r''-r'\lambda')+\frac{\dot{r}\dot{\lambda}}{r}+\ddot{\lambda}+\frac{1}{2}\dot{\lambda}^2+\frac{2\ddot{r}}{r}, \nonumber \\
& & 2\dot{r}'-\dot{\lambda}r'=0.
\label{grav6}
\end{eqnarray}
Integrating the last equation in (\ref{grav6}) gives
\begin{equation}
    e^\lambda=\frac{r'^2}{1+f(R)},
    \label{grav7}
\end{equation}
where $f$ is a function of $R$ satisfying a condition $1+f>0$ \cite{LL2}.
Substituting (\ref{grav7}) into the second equation in (\ref{grav6}) gives $2r\ddot{r}+\dot{r}^2-f=-\kappa\tilde{p}r^2$, which is integrated to 
\begin{equation}
    \dot{r}^2=f(R)+\frac{F(R)}{r}-\frac{\kappa}{r}\int\tilde{p}r^2 dr,
    \label{grav8}
\end{equation}
where $F$ is a positive function of $R$.
Substituting (\ref{grav7}) into the third equation in (\ref{grav6}) does not give a new relation.
Substituting (\ref{grav7}) into the first equation in (\ref{grav6}) and using (\ref{grav8}) gives
\begin{equation}
    \kappa(\tilde{\epsilon}+\tilde{p})=\frac{F'(R)}{r^2 r'}.
    \label{grav9}
\end{equation}
Combining (\ref{grav8}) and (\ref{grav9}) gives
\begin{equation}
    \dot{r}^2=f(R)+\frac{\kappa}{r}\int_0^R\tilde{\epsilon}r^2 r'dR.
    \label{grav10}
\end{equation}

Every particle in a collapsing fluid sphere is represented by a radial coordinate $R$ that ranges from 0 (at the center of the sphere) to $R_0$ (at the surface of the sphere).
If the mass of the sphere is $M$, then the Schwarzschild radius $r_g=2GM/c^2$ of the black hole that forms from the sphere is equal to \cite{LL2}
\begin{equation}
    r_g=\kappa\int_0^{R_0}\tilde{\epsilon}r^2 r'dR.
    \label{grav11}
\end{equation}
Equations (\ref{grav10}) and (\ref{grav11}) give
\begin{equation}
    \dot{r}^2(\tau,R_0)=f(R_0)+\frac{r_g}{r(\tau,R_0)}.
    \label{grav12}
\end{equation}
If $r_0=r(0,R_0)$ is the initial radius of the sphere and the sphere is initially at rest, then $\dot{r}(0,R_0)=0$.
Consequently, (\ref{grav12}) determines the value of $R_0$ \cite{collapse1,collapse2}:
\begin{equation}
    f(R_0)=-\frac{r_g}{r_0}.
    \label{grav13}
\end{equation}

\section{Spinless dustlike sphere}
Before considering gravitational collapse of a sphere composed of a spin fluid, it is instructive to consider spinless dust, for which the pressure vanishes and thus $\tilde{p}=0$.
Substituting (\ref{grav9}) into (\ref{grav11}) gives
\begin{equation}
    r_g=F(R_0)-F(0)=F(R_0),
    \label{dust1}
\end{equation}
which determines the value of $R_0$.
If $f<0$, then (\ref{grav8}) has a solution
\begin{equation}
    r=-\frac{F}{2f}(1+\cos\eta),\quad \tau-\tau_0(R)=\frac{F}{2(-f)^{3/2}}(\eta+\sin\eta),
\end{equation}
where $\eta$ is a parameter and $\tau_0(R)$ is a function of $R$ \cite{LL2,Tol}.
Choosing
\begin{equation}
    f(R)=-\sin^2 R,\quad F(R)=a_0\sin^3 R,\quad \tau_0(R)=\mbox{const.}
    \label{dust3}
\end{equation}
gives
\begin{equation}
    r=\frac{a_0}{2}\sin R(1+\cos\eta)\,\quad \tau-\tau_0=\frac{a_0}{2}(\eta+\sin\eta),
    \label{dust4}
\end{equation}
where $a_0$ is a constant \cite{LL2}.
Initially, at $\tau=\tau_0$ and $\eta=0$, the sphere is at rest: $\dot{r}=0$.
Clearly, a singularity $r=0$ is reached for all particles in a finite time.
The values of $a_0$ and $R_0$ can be determined from (\ref{grav13}), (\ref{dust1}), and (\ref{dust3}):
\begin{equation}
    \sin R_0=\Bigl(\frac{r_g}{r_0}\Bigr)^{1/2},\quad a_0=\Bigl(\frac{r_0^3}{r_g}\Bigr)^{1/2}.
    \label{dust5}
\end{equation}
An event horizon for the entire sphere forms when $r(\tau,R_0)=r_g$, that is, at $\cos(\eta/2)=\sin R_0$.

Substituting (\ref{dust3}) and (\ref{dust4}) into (\ref{grav7}) gives $e^{\lambda(\tau,R)}=a_0^2(1+\cos\eta)^2/4$.
If we define
\begin{equation}
    a(\tau)=\frac{a_0}{2}(1+\cos\eta),
    \label{dust6}
\end{equation}
then the square of an infinitesimal interval in the interior of a collapsing dust (\ref{grav5}) turns into \cite{LL2}
\begin{equation}
    ds^2=c^2 d\tau^2-a^2(\tau)dR^2-a^2(\tau)\sin^2 R(d\theta^2+\mbox{sin}^2\theta\,d\phi^2).
    \label{dust7}
\end{equation}
The initial value of $a$ is equal to $a_0$.
This metric has a form of the closed FLRW metric and describes a part of a closed universe with $0\le R \le R_0$.

\section{Spin-fluid sphere}
We now proceed to the main part of this Chapter and consider gravitational collapse of a sphere composed of a spin fluid.
We use the Tolman metric \cite{OS,Tol} and the EC field equations with a relativistic spin fluid as a source, which can be written as the GR field equations for a fluid source with the energy density and pressure (\ref{intro1}).
We use the temperature to represent the energy density, pressure, and fermion number density in a relativistic fluid \cite{ApJ1,ApJ2}.

Substituting $r=e^{\mu/2}$ and (\ref{grav7}) into the first equation in (\ref{grav4}) gives
\begin{equation}
    \frac{d}{d\tau}(\tilde{\epsilon}r^2 r')+\tilde{p}\frac{d}{d\tau}(r^2 r')=0,
    \label{spin1}
\end{equation}
which has a form of the first law of thermodynamics for the energy density and pressure (\ref{intro1}) \cite{ApJ1,ApJ2}.
If we assume that the spin fluid is composed by an ultrarelativistic matter in kinetic equilibrium, then $\epsilon=h_\star T^4$, $p=\epsilon/3$, and $n_\textrm{f}=h_{n\textrm{f}}T^3$, where $T$ is the temperature of the fluid, $h_\star=(\pi^2/30)(g_\textrm{b}+(7/8)g_\textrm{f})k_\textrm{B}^4/(\hbar c)^3$, and $h_{n\textrm{f}}=(\zeta(3)/\pi^2)(3/4)g_\textrm{f}k_\textrm{B}^3/(\hbar c)^3$ \cite{ApJ1,ApJ2,Gabe}.
For standard-model particles, $g_\textrm{b}=29$ and $g_\textrm{f}=90$.
Since $p'=0$, the temperature does not depend on $R$: $T=T(\tau)$.
Substituting these relations into (\ref{spin1}) gives
\begin{equation}
    r^2 r'T^3=g(R),
    \label{spin2}
\end{equation}
where $g$ is a function of $R$.
Putting this equation into (\ref{grav10}) gives
\begin{equation}
    \dot{r}^2=f(R)+\frac{\kappa}{r}(h_\star T^4-\alpha h_{n\textrm{f}}^2 T^6)\int_0^R r^2 r'dR.
    \label{spin3}
\end{equation}
Equations (\ref{spin2}) and (\ref{spin3}) give the function $r(\tau,R)$, which with (\ref{grav7}) gives $\lambda(\tau,R)$ \cite{collapse1,collapse2}.
The integration of (\ref{spin3}) also contains the initial value $\tau_0(R)$.
The metric (\ref{grav5}) depends thus on three arbitrary functions: $f(R)$, $g(R)$, and $\tau_0(R)$.

We seek a solution of (\ref{spin2}) and (\ref{spin3}) as
\begin{equation}
    f(R)=-\sin^2 R,\quad r(\tau,R)=a(\tau)\sin R,
    \label{spin4}
\end{equation}
where $a(\tau)$ is a nonnegative function of $\tau$.
This choice is analogous to a dust sphere: the first equation in (\ref{dust3}), the first equation in (\ref{dust4}), and (\ref{dust6}).
Accordingly, (\ref{spin2}) gives
\begin{equation}
    a^3 T^3\sin^2 R\cos R=g(R),
\end{equation}
in which separation of the variables $\tau$ and $R$ leads to
\begin{equation}
    g(R)=\mbox{const}\cdot \sin^2 R\cos R,\quad a^3 T^3=\mbox{const}.
\end{equation}
Consequently, we find
\begin{equation}
    aT=a_0 T_0,\quad \frac{\dot{T}}{T}+\frac{H}{c}=0,
    \label{spin7}
\end{equation}
where $a_0=a(0)$, $T_0=T(0)$, and $H=c\dot{a}/a$ is the Hubble parameter.
Substituting (\ref{spin4}) into (\ref{spin3}) gives
\begin{equation}
    \dot{a}^2+1=\frac{\kappa}{3}(h_\star T^4-\alpha h_{n\textrm{f}}^2 T^6)a^2.
    \label{spin8}
\end{equation}
Using (\ref{spin7}) in (\ref{spin8}) yields
\begin{equation}
    \dot{a}^2=-1+\frac{\kappa}{3}\Bigl(\frac{h_\star T^4_0 a^4_0}{a^2}-\frac{\alpha h_{n\textrm{f}}^2 T_0^6 a^6_0}{a^4}\Bigr).
    \label{spin9}
\end{equation}
Substituting (\ref{spin4}) into (\ref{grav7}) gives $e^{\lambda(\tau,R)}=a^2$.
Consequently, the square of an infinitesimal interval in the interior of a collapsing spin fluid (\ref{grav5}) is also given by (\ref{dust7}) \cite{collapse1,collapse2}.

The values of $a_0$ and $R_0$ can be determined from (\ref{grav13}) and (\ref{spin4}), giving (\ref{dust5}).
Substituting them and $\dot{a}(0)=0$ into (\ref{spin8}), in which the second term on the right-hand side is negligible, gives $Mc^2=(4\pi/3)r^3_0 h_\star T^4_0$.
This relation indicates the equivalence of mass and energy of a fluid sphere with radius $r_0$ and determines $T_0$.
An event horizon for the entire sphere forms when $r(\tau,R_0)=r_g$, which is equivalent to $a=(r_g r_0)^{1/2}$.
Equation (\ref{spin9}) has two turning points, $\dot{a}=0$, if \cite{Gabe}
\begin{equation}
    \frac{r^3_0}{r_g}>\frac{3\pi G\hbar^4 h_{n\textrm{f}}^4}{8h_\star^3}\sim l_\textrm{Planck}^2,
\end{equation}
which is satisfied for astrophysical systems that form black holes.

\section{Nonsingular bounce and formation of a new universe}
Equation (\ref{spin9}) can be solved analytically in terms of an elliptic integral of the second kind \cite{Gabe}, giving the function $a(\tau)$ and then $r(\tau,R)=a(\tau)\sin R$.
The value of $a$ never reaches zero because as $a$ decreases, the right-hand side of (\ref{spin9}) becomes negative, contradicting the left-hand side.
The change of the sign occurs when $a<(r_g r_0)^{1/2}$, that is, after the event horizon forms.
Consequently, all particles with $R>0$ fall within the event horizon but never reach $r=0$ (the only particle at the center is the particle that is initially at the center, with $R=0$).
A singularity is therefore avoided and replaced with a regular bounce \cite{collapse1,collapse2}.
Nonzero values of $a$ in (\ref{dust7}) give finite values of $T$ and therefore finite values of $\epsilon$, $p$, and $n_\textrm{f}$.

If the initial mass of a spin-fluid sphere is insufficient to form an event horizon, then the fluid bounces and disperses back to the region of space outside the sphere \cite{Iran}.
When an event horizon forms, the fluid cannot disperse back to the region of space outside the horizon because of the unidirectionality of the motion of matter through a horizon \cite{LL2}.
Moreover, it cannot tend to a static state because the spacetime within an event horizon is not stationary.
Consequently, the spin fluid on the other side of the event horizon must expand as a new, growing universe with a closed geometry (constant positive curvature) \cite{cosmo}.
This universe can be regarded as the three-dimensional surface of a four-dimensional sphere with radius $a(\tau)$, which is the scale factor of this universe.
The new, closed universe is oscillatory: the value of $a$ oscillates between the two turning points.
The value of $R_0$ does not change.
A turning point at which $\ddot{a}>0$ is a bounce, and a turning point at which $\ddot{a}<0$ is a crunch.
The universe has therefore an infinite number of bounces and crunches, and each cycle is alike.

The Raychaudhuri equation for a congruence of geodesics without rotation and four-acceleration is $d\theta/ds=-\theta^2/3-2\sigma^2-P_{ik}u^i u^k$, 
where $\theta$ is the expansion scalar and $\sigma^2$ is the shear scalar \cite{Niko}.
For a spin fluid, the last term in this equation is equal to $-\kappa(\tilde{\epsilon}+3\tilde{p})/2$.
Consequently, the necessary and sufficient condition for avoiding a singularity in a black hole is $-\kappa(\tilde{\epsilon}+3\tilde{p})/2>2\sigma^2$.
For a relativistic spin fluid, $p=\epsilon/3$, this condition is equivalent to
\begin{equation}
    2\kappa\alpha n_\textrm{f}^2>2\sigma^2+\kappa\epsilon.
    \label{avoid1}
\end{equation}
Without torsion, the left-hand side of (\ref{avoid1}) would be absent and this inequality could not be satisfied, resulting in a singularity.
Torsion therefore provides a necessary condition for preventing a singularity.
In the absence of shear, this condition is also sufficient.

The presence of shear opposes the effects of torsion.
The shear scalar $\sigma^2$ grows with decreasing $a$ like $\sim a^{-6}$, which is the same power law as that for $n_\textrm{f}^2$ \cite{Kop1,Kop2}.
Therefore, if the initial shear term dominates over the initial torsion term in (\ref{avoid1}), then it will dominate at later times during contraction and a singularity will form.
To avoid a singularity if the shear is present, $n_\textrm{f}^2$ must grow faster than $\sim a^{-6}$.
Consequently, fermions must be produced in a black hole during contraction.

\section{Particle production}
The production rate of particles in a contracting or expanding universe \cite{prod7,prod1,prod2,prod4,prod5,prod3,prod6} can be phenomenologically given by
\begin{equation}
    \frac{1}{c\sqrt{-g}}\frac{d(\sqrt{-g}n_\textrm{f})}{dt}=\frac{\beta H^4}{c^4},
    \label{part1}
\end{equation}
where $g=-a^6\sin^4R\sin^2\theta$ is the determinant of the metric tensor in (\ref{dust7}) and $\beta$ is a nondimensional production rate \cite{ApJ1,ApJ2}.
With particle production, the second equation in (\ref{spin7}) turns into
\begin{equation}
    \frac{\dot{T}}{T}=\frac{H}{c}\Bigl(\frac{\beta H^3}{3c^3 h_{n\textrm{f}}T^3}-1\Bigr).
    \label{part2}
\end{equation}
Particle production changes the power law $n_\textrm{f}(a)$:
\begin{equation}
    n_\textrm{f}\sim a^{-(3+\delta)},
\end{equation}
where $\delta$ varies with $\tau$ \cite{collapse1,collapse2}.
Putting this relation into (\ref{part1}) gives
\begin{equation}
    \delta\sim -a^\delta\dot{a}^3.
\end{equation}

During contraction, $\dot{a}<0$ and thus $\delta>0$.
The term $n_\textrm{f}^2\sim a^{-6-2\delta}$ grows faster than $\sigma^2\sim a^{-6}$ and a singularity is avoided \cite{collapse1,collapse2}.
Particle production and torsion act together to reverse the effects of shear, generating a nonsingular bounce.
The dynamics of the nonsingular, relativistic universe in a black hole is described by equations (\ref{spin8}) and (\ref{part2}),
with the initial conditions $a(0)=(r_0^3/r_g)^{1/2}$ and $\dot{a}(0)=0$, that give the functions $a(\tau)$ and $T(\tau)$.
The shear would enter the right-hand side of (\ref{spin8}) as an additional positive term that is proportional to $a^{-4}$.
When the universe becomes nonrelativistic, the term $h_\star T^4$ in (\ref{spin8}) changes into a positive term that is proportional to $a^{-1}$.
The cosmological constant enters (\ref{spin8}) as a positive term that is proportional to $a^{2}$.

Particle production increases the maximum size of the scale factor that is reached at a crunch.
Consequently, the new cycle is larger and lasts longer then the previous cycle.
According to (\ref{dust5}), $R_0$ is given by
\begin{equation}
    \sin^3 R_0=\frac{r_g}{a(0)},
\end{equation}
where $a(0)$ is the initial scale factor that is equal to the maximum scale factor in the first cycle.
Since the maximum scale factor in the next cycle is larger, the value of $\sin R_0$ decreases.
As cycles proceed, $R_0$ approaches $\pi$.

Without torsion and particle production, a singularity would be reached and the metric would be described by the interior Schwarzschild solution, which is equivalent to the Kantowski--Sachs metric describing an anisotropic universe with topology $R\times S^2$ \cite{KS2,KS1,KS3}.
Thanks to torsion, the universe in a black hole becomes closed with topology $S^3$ (3-sphere).

\section{Inflation and oscillations}
During contraction, $H$ is negative and the temperature $T$ increases.
During expansion, if $\beta$ is too big, then the right-hand side of (\ref{part2}) could become positive.
In this case, the temperature would grow with increasing $a$, which would lead to eternal inflation \cite{ApJ1,ApJ2}.
Consequently, there is an upper limit to the production rate: the maximum of the function $(\beta H^3)/(3c^3 h_{n\textrm{f}}T^3)$ must be lesser than 1.

If $(\beta H^3)/(3c^3 h_{n\textrm{f}}T^3)$ in (\ref{part2}) increases after a bounce to a value that is slightly lesser than 1, then $T$ would become approximately constant.
Accordingly, $H$ would be also nearly constant and the scale factor $a$ would grow exponentially, generating inflation.
Since the energy density would be also nearly constant, the universe would produce enormous amounts of matter and entropy.
Such an expansion would last until the right-hand side of (\ref{part2}) drops below 1.
Consequently, inflation would last a finite period of time.
After this period, the effects of torsion weaken and the universe smoothly enters the radiation-dominated expansion, followed by the matter-dominated expansion.

If the universe during expansion does not reach a critical size at which the cosmological constant is significant, then it recollapses to another bounce and starts a new oscillation cycle \cite{cc1,cc2}.
The new cycle is larger and longer then the previous cycle \cite{ent1,ent2,ApJ1,ApJ2}.
After a finite series of cycles, the universe reaches the critical size which prevents the next contraction and enters the cosmological-constant-dominated expansion, during which it expands indefinitely.
The value of $R_0$ asymptotically tends to $\pi$, which is the maximum value of $R$ in a closed isotropic universe given by (\ref{dust7}).
The last bounce, referred to as the big bounce, is the big bang.

A more realistic scenario of gravitational collapse should involve a fluid sphere that is inhomogeneous and rotating.
If the pressure in the sphere is not homogeneous, then the system of coordinates cannot be comoving and synchronous \cite{LL2,Lif}.
Consequently, $\nu$ and the temperature would depend on $R$ and the equations of the collapse and the subsequent dynamics of the universe would be more complicated.
If the sphere were rotating, then further complications would appear \cite{Dor} and the angular momentum of the forming Kerr black hole would be another parameter in addition to the mass \cite{Kerr}.
Nevertheless, the general character of the effects of torsion and particle production in avoiding a singularity and generating a bounce in a black hole would still be valid.

Torsion and particle production act together to reverse gravitational attraction generated by shear and prevent a singularity, to turn the interior of a black hole into a new universe, and to generate inflation in that universe \cite{collapse1,collapse2}.
Torsion may also explain the matter-antimatter asymmetry in the universe \cite{anti}.
In addition, it could explain the cosmological constant, which is necessary for a closed universe to expand to infinity \cite{exp}.
Furthermore, torsion may impose a spatial extension of fermions \cite{non} and eliminate the ultraviolet divergence of radiative corrections represented by loop Feynman diagrams in quantum field theory \cite{toreg}.

If every black hole becomes an Einstein--Rosen bridge to a new universe on the other side of its event horizon \cite{Ein,ER}, then our universe might have been born as a baby universe in a parent black hole existing in another universe.
This hypothesis, following from the presented analysis of gravitational collapse of a spin fluid \cite{collapse1,collapse2}, naturally solves the black hole information paradox: the information about the initial state of a collapsing matter is not lost but it goes through the event horizon to the new universe \cite{cosmo}.
Furthermore, inflation generated by torsion and particle production, is consistent with the Planck observations of the cosmic microwave background radiation \cite{SD}.\\

\noindent
I am grateful to Francisco Guedes and my Parents, Bo\.{z}enna Pop{\l}awska and Janusz Pop{\l}awski, for inspiring this work.

\end{document}